\documentclass[referee]{raa}
\usepackage{graphicx,times}

\begin{document}

\title{A Monte Carlo Study of the Spectra from Inhomogeneous Accretion Flow}

   \volnopage{Vol.0 (200x) No.0, 000--000}      
   \setcounter{page}{1}          

   \author{Xiao-Di Yu\inst{1,3}
   \and Ya-Ping Li\inst{2,3}
   \and Fu-Guo Xie\inst{2,3}
   \and Ren-Yi Ma\inst{1,3}}

 \institute{Department of Astronomy and Institute of Theoretical Physics and Astrophysics, Xiamen University, Xiamen, Fujian 361005, China; {\it ryma@xmu.edu.cn} \\
   \and Key Laboratory for Research in Galaxies and Cosmology, Shanghai Astronomical Observatory, Chinese Academy of Sciences, 80 Nandan Road, Shanghai 200030, China \\
   \and SHAO-XMU Joint Center for Astrophysics, Xiamen, Fujian 361005, China}

\date{Received~~2009 month day; accepted~~2009~~month day}

\abstract {The model of inhomogeneous accretion flow, in which cold clumps are surrounded by hot gas or corona, has been proposed to explain the spectral features of black hole X-ray binaries (BHXBs).
In this work, we try to find possible observational features in the continuum that can indicate the existence of clumps.
The spectra of inhomogeneous accretion flow are calculated via the Monte Carlo method. Since the corresponding accretion flow is unsteady and complex,
the accretion flow is described by a set of free parameters, the ranges of which can include the real cases.
The influences of the parameters are investigated.
It is found that the thermal component of the spectra deviates from the multi-color black body spectra in the middle power-law part.
On the one hand, a warp appears due to the gap region between the clumps and the outer cold disk, and on the other hand, the slope of the line connecting the thermal peaks deviates from 4/3.
The warp feature, as well as the correlation between the thermal peak at higher frequency and the spectral index, are possible to indicate the existence of clumps, and are worthy of further investigation with more self-consistent models.
\keywords{accretion, accretion disks -- X-rays: binaries -- stars: black holes}
}

   \authorrunning{X.-D. Yu et al. }
   \titlerunning{The Spectra of Inhomogeneous Accretion Flow}

\maketitle

\section{Introduction}\label{intro}

It has been widely accepted that both the soft and hard states of BHXBs can be explained by the model of hybrid accreting flow,
i.e., the standard accretion disk (SAD, Shakura \& Sunyaev~\cite{SS73}) plus optically thin hot gas such as the corona or the advection dominated accretion flow (e.g., Narayan \& Yi~\cite{NY94}; Yuan \& Narayan~\cite{YN14} and references therein).
In the soft state, the SAD extends to the innermost stable circular orbit (ISCO) of the black hole, being sandwiched by weak corona (e.g., Liang \& Price~\cite{LP77}; Zhang et al.~\cite{Zhang97}).
While in the hard state, the SAD is truncated by the hot accretion flow at a certain radius about several tens to one hundred of gravitational radius, $R_{\rm g}\equiv GM/c^2$ (e.g., Yuan \& Narayan~\cite{YN14}).
Given this hybrid model, the state transition corresponds to the variation of hot gas,
and more importantly, the transformation of the cold disk between the truncated SAD and the complete SAD that extends to the ISCO.
However, the detailed process of transformation of the cold disk remains unclear.

The most natural way is the gradual extending or receding of the truncation radius,
as first proposed by (Esin et al.~\cite{E97}).
However, this model is difficult to explain the hysteresis phenomenon, i.e. the hard-to-soft state transition during the rise phase
occurs at a higher luminosity than the soft-to-hard one during the decline phase in a single outburst (e.g., Zdziarski \& Gierli{\'n}ski~\cite{ZG04}; Done et al.~\cite{D07}).
The hysteresis indicates that other factors, besides the dimensionless accretion rate, play roles during state transitions.
Different scenarios have been proposed, for example,
the different Compton cooling on the corona for different accretion modes (e.g., Meyer-Hofmeister et al.~\cite{M05}; Liu et al.~\cite{LMM05}),
the rather different accretion flow formed by accreting low angular momentum material from the stellar wind (e.g., Smith et al.~\cite{S02}; Maccarone \& Coppi~\cite{MC03}), the dramatic changes in accretion flow during the H ionization instability (Done~\cite{D07}), the magnetic processes (e.g., Petrucci et al.~\cite{P08}; Balbus \& Henri~\cite{BH08}; Begelman \& Armitage~\cite{BA14}),
and the outflow (Cao~\cite{Cao16}).

However, some studies have shown that the transformation of the cold disk could be more complicated.
Instead of continuous cold disk, an interrupted cold disk can form, with an inner cold disk separating from the outer cold disk by a gap of hot gas.
Considering the relative rate of evaporation and condensation,
theoretical works have shown that the cold disk in the inner region
forms earlier but disappear later
than that in the middle region (e.g., Meyer et al.~\cite{M07}; Liu et al.~\cite{L07}, \cite{L11}; Qiao \& Liu~\cite{QL12}).
Therefore, it is possible to form an interrupted cold disk
when the accretion rate is slightly lower than the rate during state transition.

In fact the geometry of the cold disk could be even more complicated.
Instead of continuous inner recondensed cold disk, cold clumps are possible to form in the inner region for different reasons.
Firstly, the instabilities of the radiation-dominated regions, such as thermal, magnetorotational and photon-bubble instabilities,
induce cold clumps in the accretion flow (e.g., Krolik~\cite{K98}; Gammie~\cite{G98}; Begelman~\cite{B01}; Blaes \& Socrates~\cite{BS03}).
Secondly, overcooling can also produce cold clumps.
As accretion rate increases, hot accretion flow can only exist beyond a
certain radius, within which the Coulomb energy transfer is so efficient
that it is stronger than the sum of
the viscous dissipation and compression work (Yuan~\cite{Y01}, \cite{Y03}).
The only viable solution is the cold optically thick disk, which means that
the hot accretion flow collapses onto the equatorial plane and cold clumps form (Xie \& Yuan~\cite{X12}; Yang et al.~\cite{Y15}).
Numerical simulations have shown such cold clumps clearly (Barai et al.~\cite{BPN12}; Wu et al.~\cite{W16}; Sadowski et al.~\cite{Sadowski17}).

Such inhomogeneous accretion flow has been proposed a long time before
to explain the spectral features in AGNs (Kuncic et al.~\cite{K96}).
The spectral and timing properties of the inhomogeneous accretion flow have been studied by many author (e.g., Malzac \& Celotti~\cite{MC02}; McClintock et al.~\cite{MMFR06}; Dexter \& Quataert~\cite{DQ12}).
Recently, Yang et al.~(\cite{Y15}) found that such model can explain the correlation between the photon-index and luminosity for both BHXBs and AGNs successfully.

Since the accretion flow is unsteady in the transition stage,
it is difficult to investigate the physical state of the accretion flow.
Wang et al.~(\cite{W12}) investigated the dynamics of inhomogeneous accretion flow by solving the Boltzmann equations, but further studies are still needed.
Numerical simulations should be a good way, but it is still difficult and time consuming.
It would be helpful if the physical conditions can be better constrained from observations.
In this paper, as a first step, the physical condition of the inhomogeneous accretion flow is described by a set of free parameters, and the influences of these parameters on the continuum spectra are investigated in order to explore possible observable evidences for the existence of clumps.
Since the geometry of the accretion flow is complex, it is impossible to
calculate the spectra analytically, so the Monte Carlo method is taken in our work.

In Sec.2 the model and parameters are introduced and in Sec.3 the results are shown. Discussion and summary are given in Sec.4 and Sec.5, respectively.

\section{Toy Model}\label{toy}

\begin{figure}
   \centering
  \includegraphics[width=8cm, angle=0]{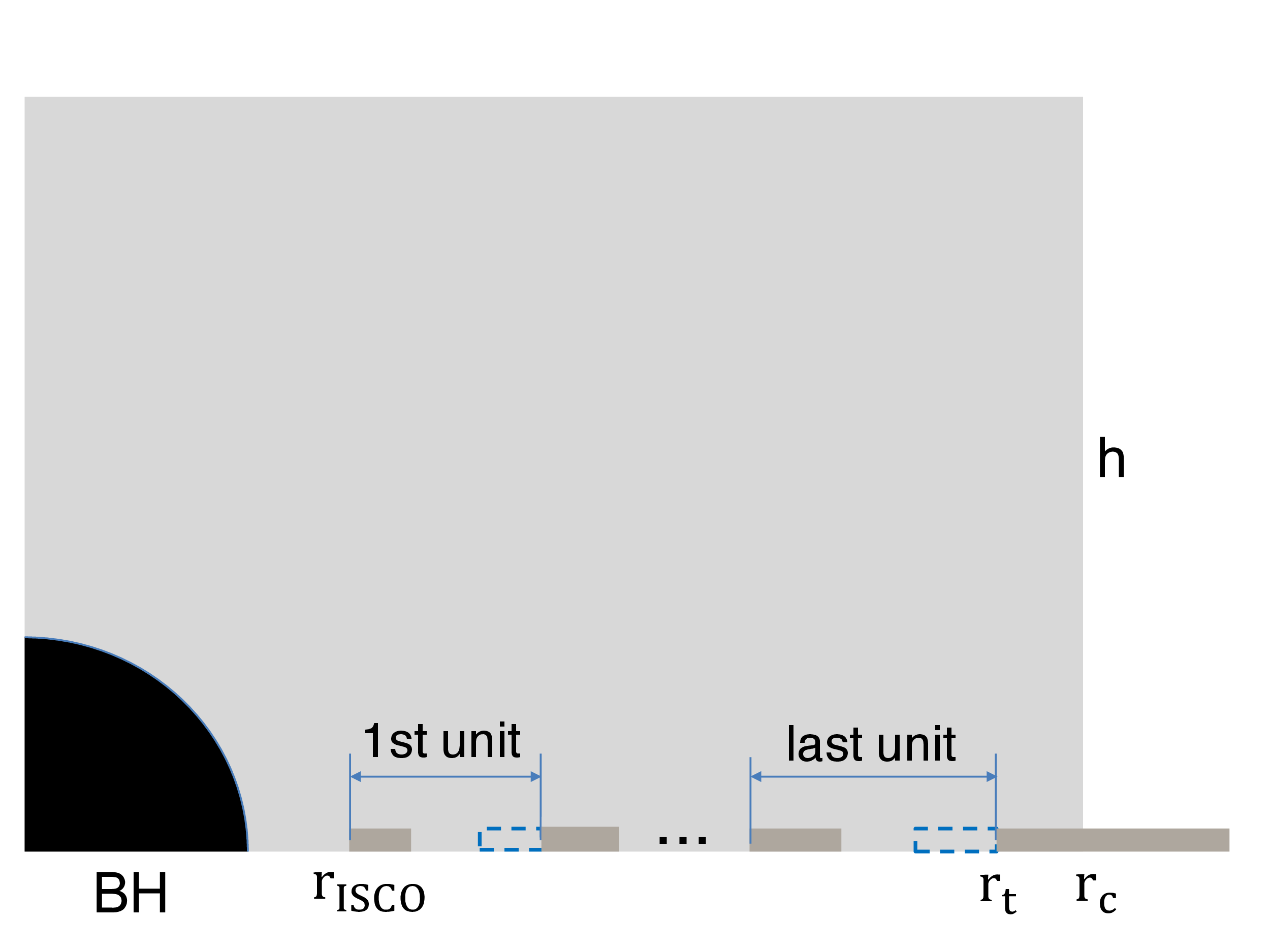}
\caption{One-quarter cross-section schematic diagram of the 3D toy model for the case $f>0$. The cold disk and clumps are represented by dark gray rectangle, the hot phase flow is the region filled with light gray, and the black hole is shown with black.
The dashed rectangles show the position of clumps when $f<0$.
The parameters $r_{\rm ISCO}$, $r_{\rm_{t}}$, $r_{\rm_{c}}$ and $h$ correspond to the radius of ISCO, the radius of truncation, the radial range and the height of hot phase flow, respectively. }
\label{fig:syt}
\end{figure}

\subsection{Parameters of The Hot Accretion Flow}\label{res.c}
\begin{figure*}
   \centering
  \includegraphics[width=18cm, angle=0]{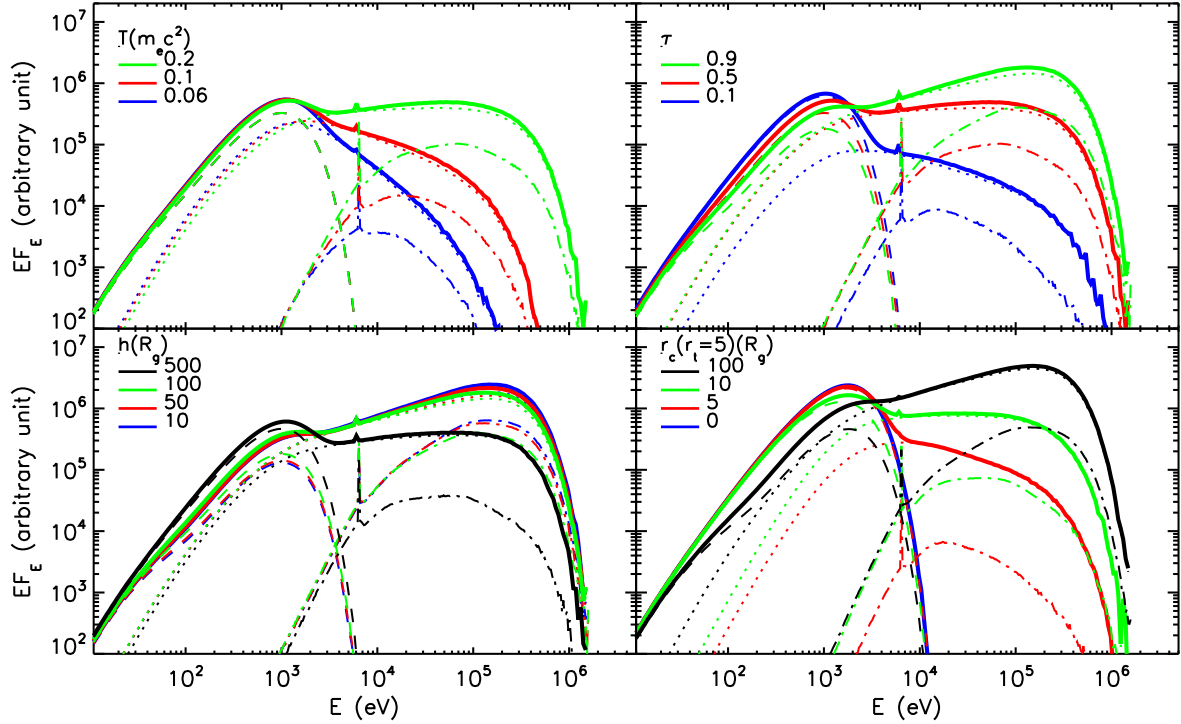}
\caption{This figure shows the effects of parameters of corona on the output spectrum.
The dashed, dotted, and dot-dashed lines show the components of thermal radiation (un-scattered photons) of the cold disk, the power-law tail and the reflection, respectively, and the solid lines show the total spectra. Different colors represent different parameters.}
\label{fig:c}
\end{figure*}

Our model is basically a truncated disk model (Yuan \& Narayan~\cite{YN14} and references therein), i.e., there is a cold SAD outside certain radius, within which it is replaced by hot gas (the advection dominated accretion flow,
the luminous hot accretion flow or the corona).
The difference is that some cold clumps are distributed in the hot accretion flow/corona.
For simplicity, we assume the clumps and the outer cold disk to lie on the equatorial plane.
The clumps are represented by a series of concentric annuli, because
the clumps should take the form of long arc due to the tidal stretching force and the differential rotation of the accretion flow.
Without the knowledge of the size of the clumps, we assume them to be proportional to their radial position, i.e., evenly distributed in the space of logarithmic radius.

Such a model can be described by a set of parameters.
For the corona, there are four parameters, the temperature of hot electrons $T$,
the vertical optical depth $\tau$,
the height $h$, and the radial range $r_{\rm_{c}}$.
In order to investigate the influences of the parameters,
we assume the corona to be slab-like and homogeneous in density and temperature.
For the cold disk, we use three parameters,
the truncation radius $r_{\rm_{t}}$ where the outer cold disk is truncated,
the number of clumps $n$,
and the covering factor $f$, which describes the ratio of the total width of all the clumps to that of the unstable region, or the region inside the truncation radius.
To describe the position of the clumps, the unstable region is separated into several units.
Each unit is made up of one clump and one neighbouring gap.
Considering that the clumps can occupy either the inner or outer part of the unit,
in order to tell the difference, the covering factor $f$ is set to be positive or negative, respectively.
Therefore the value of $f$ is in the range [-1,1].
For $f=0$, it corresponds to the case without clumps inside $r_{\rm_{t}}$;
for $f=\pm 1$, it corresponds to the cases that the gaps disappear and the cold disk continuously extends to the ISCO;
and if $n=1$, for $0<f<1$, it corresponds to the model predicted by the recondensation of the corona (e.g., Liu et al.~\cite{L11}).

The spectra of such inhomogeneous accretion flow can be well calculated with the Monte Carlo approach (e.g., Pozdnyakov~\cite{P83}; Ma et al.~\cite{Ma06}),
which follows a large amount of seed photons,
randomly draws their interactions with the electrons, until the photons finally escape from the system.
The seed photons come from the thermal radiation of the cold clumps or disk,
and then pass through the corona directly or deviate from the initial direction
because of the scattering by the electrons in corona.
The scattered photons can encounter the cold disk, cross the equatorial plane through the gap,
or travel upward like the seed photons.
If encounter the cold disk,
photons can be absorbed due to photon-ionization or scattered by electrons in the cold disk (e.g., Done~\cite{D10}; Fabian \& Ross~\cite{FR10}).
A consequence of the ionization is the emission of recombination lines.
The reflected photons from the cold disk will cross the corona once again like the seed photons.
The seed photons can be followed until they escape from the corona or lose energy completely without line emission.
Sum up all the escaped photons, and the spectrum can be obtained.
The total spectrum consists of three components,
the thermal component from the seed photons that directly escape without being scattered,
the power-law component produced by Compton scattering in the corona,
and the reflection component by the cold disk and clumps.

We note that the hot accretion follow can also emit by synchrotron radiation, as it is magnetized and the electrons are mildly relativistic.
Therefore, the synchrotron emission should also serve as seed photons for the Compton scattering. However, since we concentrate on the transition stage, in which the thermal radiation dominates over the synchrotron radiation,
the synchrotron seed photons are ignored in this paper.

As the accretion flow is unsteady and the physical processes are still not
well understood, we set the parameters free so that the actual processes can be included
in the parameter space.
The physical limitations can be roughly taken into account by adopting reasonable parameters.
For example, when the number of soft photons increases, considering the energy balance between the clumps and hot gas, relatively small optical depth or
temperature should be taken for the corona.

\section{Results}\label{res}

\begin{figure*}
   \centering
  \includegraphics[width=18cm, angle=0]{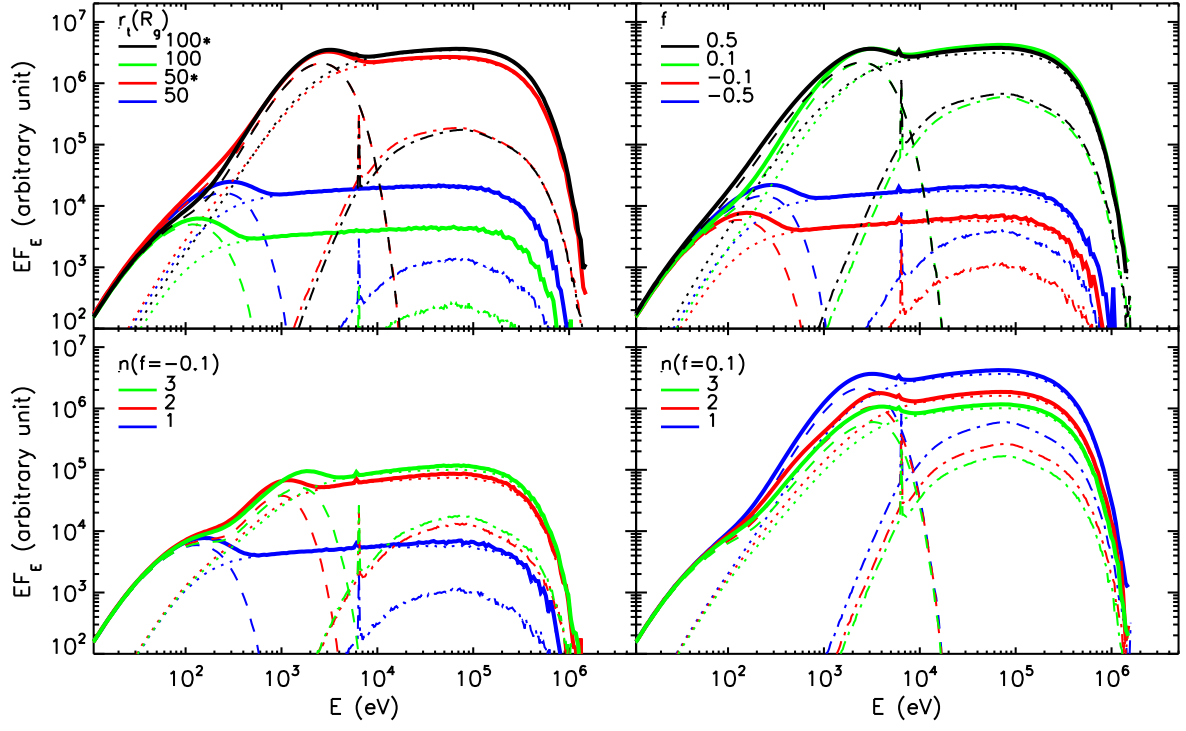}
\caption{This figure shows the effects of parameters of cold disk on the output spectrum.
The line types are the same as those of Figure~\ref{fig:c}.
The lines marked with and without '*' in the upper left panel correspond to the cases with and without a inner clump. For comparison, the inner clump is set to cover the region from $R_{\rm ISCO}$ to $10 \mbox{ } {R_{\rm_{g}}}$.
}
\label{fig:d}
\end{figure*}

In this section, we show the output spectra for different sets of parameters.
As default we take the parameters as their typical value, i.e., $r_{\rm_{t}}=100\mbox{ } {R_{\rm_{g}}}$, $f= 0.1$, $n= 1$, $kT= 0.2 \mbox{ } {m_{\rm_{e}}c^{2}}$, $\tau = 0.5$, $h=100\mbox{ } {R_{\rm_{g}}}$, $r_{\rm_{c}}= 200\mbox{ } {R_{\rm_{g}}}$.
The outer cold disk is fixed at 1000$R_{\rm_g}$.
The mass of the black hole is taken as $10 M_{\odot}$.
The influences of these parameters are shown in Figures~\ref{fig:c} and \ref{fig:d}.
The solid lines show the total spectrum.
The dashed, dotted, and dot-dashed lines show the thermal, power-law and reflection components, respectively.
Different colors represent different sets of parameters.

\subsection{Parameters of The Hot Gas}
The influences of the corona temperature and optical depth are clearly shown in the upper two panels of Figure~\ref{fig:c}.
Both parameters can significantly affect the hardness of the spectrum.
The corona temperature is important to determine the high-energy cutoff, and irrelevant to the thermal component of the spectrum.
While the optical depth is irrelevant to the energy cutoff, but important to the thermal component.
The result is natural because on the one hand, the energy that scattered photons saturated is only determined by electrons temperature $T$, i.e. about $4kT$, and on the other hand the escaping probability of soft seed photons is proportional to $e^{-\tau}$.

For larger optical depth, the slope of thermal component in the middle power-law part about 0.1-1 keV, which is roughly a line connecting the peaks of black-body spectra at different radii, deviates further from  that of continuous SAD, being flatter than 4/3.
At lower frequencies, the radiation is dominated by the outer disk,
where the averaged optical depth of the corona is smaller and most photons can directly escape.
However, at higher frequencies, the radiation are dominated by the inner disk, where
the averaged optical depth of the corona over different direction is larger, and relatively less photons can escape.
As a result, the power-law index of the thermal component becomes smaller than 4/3.

The lower left panel of Figure~\ref{fig:c} shows that the trend of the output spectra for various $h$ is similar to that of $\tau$. The reason is as follows.
When $h$ increases, the optical depth per unit length decreases for given vertical optical depth.
As the radial range of the corona remains the same, the averaged optical depth decreases, and the slope of the power-law component becomes flatter.

The radial scale of the corona is important as shown in the lower-right panel Figure~\ref{fig:c}.
It can affect both the thermal component and the power-law tail.
The power-law component is harder for larger $r_{\rm_{c}}$,
while the thermal component is similar to multi-color black body but with index smaller than 4/3.
Both effects arise from the larger averaged optical depth of the corona for larger $r_{\rm_{c}}$.

\subsection{Parameters of The Clumps}\label{res.d}

In the upper left panel of Figure~\ref{fig:d}, we show the influences of truncation radius for the cases without cold clumps and the cases with one clump  in the innermost region.
Firstly, it can be seen that the truncation radius is important if there is no cold clump inside the truncation radius,
but when cold clumps exist in the innermost region, it is not important.
In the former case,
all the soft seed photons are from the outer cold disk.
Since the disk temperature and flux are higher at smaller radius,
when the truncation radius extends to smaller radius,
the peak flux and peak frequency of the thermal component, as well as
the final flux, all become higher.
While in the latter case, the seed photons are dominated by the clumps in the
inner region,
and so the truncation radius is not important any more.
Secondly, since the inner disk disconnects with the outer disk, the middle part of the thermal spectrum
is not power-law any more, taken instead by a warp between the thermal peaks of the inner and outer cold disks.
Thirdly, given the inner clumps, the larger truncation radius means larger gap and higher probability for photons to cross the equatorial plane or larger averaged optical depth.
Consequently, the reflection component decreases, but the power-law component becomes slightly harder.

The influences of the covering factor is shown in the upper right panel.
For smaller covering factor, the clumps are smaller and the gaps between clumps are larger.
As the curves show, the influences of the clump size depend on the clump position.
If the clump lies in the innermost region, when $f$ increases from 0.1 to 0.5,
the flux of the final spectrum does not changes significantly,
while if the clump lies close to the truncation radius, similar to the cases
with different truncation radius, the final spectrum changes significantly with the clump size.
This is also because the thermal radiation of the clumps decrease with increasing radius.
Additionally, for larger $f$, the gap region is smaller, the possibility
for photons to cross the equatorial plane or the averaged optical depth reduces,
and the power-law component becomes slightly softer.
Considering the evolution of the clumps, $f$ could vary between 0 to 1.
If $f$ approaches 1, the result would be similar to the case $f=0.5$, but with steeper hard X-ray spectra.
If the clumps are tiny, or $f$ is very small and positive, the thermal flux becomes dependent on the clump size as will shown in next paragraph.

The effects of the number of cold clumps can be seen from the lower two panels.
The thermal component varies for different number of clumps.
For a given negative $f$, larger number of cold clumps means smaller width of the clumpy units,
the clumps can therefore approach the ISCO closer, and so the luminosity and peak frequency are higher.
While for a given positive $f$, more cold clumps mean less cold disk in the innermost region and therefore lower luminosity.
No matter $f$ is positive or negative,
when $n$ is large enough the gaps will become so small that the warp feature disappears,
and the middle part of thermal component becomes power-law again just as that of the continuous disk,
but the slope is smaller since the clumps only cover part of the equatorial plane.
Moreover, for large $n$, the clumpy units are very small, and it can be expected that the sign of $f$ does not affect the spectra much. 

Combining the upper right panel and the lower right panel of figure~\ref{fig:d},
it can be noticed that the flux of the thermal peak is not sensitive to the clump
size in the upper panel, but in the lower panel, the flux of the thermal peak
changes with the number or the size of the clumps.
This is because of the assumption that the temperature of clumps follow that of SAD,
in which there is a peak for the disk temperature at about several $R{\rm_g}$,
which is a little larger than the radius of ISCO.
If the innermost clump is large enough to cover the region of highest temperature,
the increase of its size will not contribute much to the peak flux.
But if only part of the hottest region is covered, the flux depends on the clump size.
No matter how much the peak flux is, the peak frequency is almost the same.
This is because the clumps can always cover more or less the hottest region.

In general, the clumps have three influences on the spectra.
Firstly, the thermal peak at higher frequency depends on the position and the size
of the hottest clump.
Secondly, instead of a power-law form, a warp appears in the middle part of the thermal component since the gaps between the clumps do not contribute to the thermal emission.
Moreover, the covering factor can also influence the power-law component from Compton scattering by changing the averaged optical depth or the probability for photons to cross the equatorial plane, but this effect is of minor importance.

\section{Discussion}\label{dis}

Possible observable features of the clumps are what we concern in this paper.
The power-law component of the total spectrum by Compton scattering is not viable to detect clumps because it is almost completely determined by the state of the corona.
In our calculations, since the clumps are assumed to be neutral, the reflection component is also not helpful to detect the clumps.
However, since the clumps are surrounded by the hot gas, it is possible that the
ionization degree is greatly different to that of SAD.
In this case, some clues may be found from the reflection continuum.
It can also be expected that general relativistic effects are different at different radius,
so the profile of relativistic Fe K$\alpha$ line may be different if the clumps exist.
But these are out of the scope of this paper.
Due to the simplicity of the model, we concentrate on the thermal component.

As mentioned in last section, there are three features for the thermal component,
i.e., the thermal peak of higher frequency,
the warp between thermal peaks,
and the slope of the line connecting the thermal peaks.
The thermal peak of higher frequency depends on the inner edge of the outer cold disk or the clump of highest temperature.
So a single such peak is not enough to show the clumps.
But if the peaks at different moment during transition are given,
clues of the clumps may be found.
The warp is also possible to indicate the clumps because it is only determined by the gaps between the clumps and outer cold disk.
The slope of the line connecting the thermal peaks cannot be used,
because it can be ascribed to the averaged effective optical depth of the corona.

\subsection{Correlation Between Thermal Peak And Spectral Index}
When the hard-to-soft state transition occurs, the X-ray flux does not vary much,
but the shape of the spectral, or the spectral index, changes significantly.
This implies the conversion between the hot phase and cold phase.
As the hot phase gas condenses into cold phase, the hard X-ray photons decrease.
Meanwhile, the soft thermal X-ray photons increase as the size of the clumps increases and more seed photons are emitted.
So it is possible for the radiative efficiency to remain the same (Xie \& Yuan~ \cite{X12,X16}).

Considering our results, as shown in the upper right and lower right panels of Figure~\ref{fig:d},
we expect that the detailed process of the transition from truncated SAD to continuous SAD could be identified from the correlation between the thermal peak and the spectral index.
If the truncation radius gradually contracts during state transition, the peak flux and peak frequency keep increasing with the spectral index.
If an inner cold disk forms first and expands gradually to the outer cold disk,
although the flux of the thermal component increases but the peak flux and peak frequency remain the same.
So the peak flux and peak frequency do not vary with the spectral index.
If many clumps appear simultaneously at the beginning of transition,
the peak flux increases when the clumps expand, but the peak frequency remain
almost the same.
Therefore, the peak flux is positively correlated with the spectral index,
but the peak frequency is uncorrelated.

If we consider the possible limitations from the energy interactions between the corona and clumps or cold disk,
the temperature of the clumps and inner edge of the outer cold disk
should be higher than that of SAD at the beginning, but later both are close to that of SAD.
In this case, for the model of clumps, the possible highest clump temperature decreases,
while for the model of extracting truncation radius, the temperature of
the inner edge of the cold disk will not increases as much as the cases without the heating from corona.
So the correlation index of the positive correlations mentioned above should decrease,
and the uncorrelated relations may become anti-correlated.
Further investigations based on more consistent model are needed to give the
correlation indexes.

\subsection{The Warp Feature In The Thermal Component}
The degree of warp depends on the size and temperature of the clumps.
As for our model, with the clump temperature being assumed to be that of SAD at the same radius, the warp depends on the gap.
If the gap is larger, the warp would be more significant,
which can be seen from the lower right panel of Figure~\ref{fig:d}.

During the hard-to-soft transition, the gaps between the clumps and outer cold
disk become smaller, and the warp becomes less obvious.
So we can not find the warp at the end of the transition.
Similarly, it is also difficult to find the warp at the begin of the soft-to-hard transition.

The warp structure may become even less obvious if more realistic soft seed photons or relativistically blurred emission lines are taken.
On the one hand, for the temperature we concerned, the bound-free opacity and bound-bound opacity can modify
the spectrum, leading to a notable deviation from the color-corrected black body (Shimura \& Takahara~\cite{ST95}).
On the other hand, the relativistically blurred emission lines due to Fe L shell,
C, N, and O may also fall into this energy range.
These reflection features could be significant and have been proposed to explain the soft X-ray excess in AGN (e.g., Crummy et al.~\cite{Crummyetal06}).
Further study are needed to investigate these effects.

The heating effect of the outer cold disk by the hard X-rays can also produce similar warp shape (e.g., Gierli{\'n}ski et al.~\cite{GDP08}; Chiang et al.~\cite{Chiang10}).
One possible way to identify the origin is the flux correlation between the X-ray and UV bands.
For reprocessing model, if X-ray flux decreases, the UV bump will also
decrease after  a certain time lag,
but for the inhomogeneous flow,
the correlation between UV bump and the X-ray flux is complex and case dependent.
For example,
when the accretion rate increases, the UV bump can increase together with X-ray flux;
If the accretion rate remain the same,
they can be either anti-correlated due to energy balance between the corona and clumps, or
none-correlated due to energy injection into the corona such as the dissipation of magnetic field.

Additionally, radiation in the UV/soft X-ray band suffers strong attenuation by the interstellar absorption, so carefully designed and high-resolution observations are needed in order to detect the warp feature.

\section{Conclusions}
\label{sec:conc}

In order to explore possible detectable signals of the clumps in the inhomogeneous accretion flow,
we parameterize the model and investigate the influences of each parameter,
so that the actual process can be inferred from the variation of parameters.
We find that the clues of the clumps may be found in the warp feature of the middle part of the thermal component,
as well as the correlation between the thermal peak at higher frequency and the spectral index.

More self-consistent models, which includes the more realistic disk spectrum, the synchrotron
seed photons, the interaction between the clumps and disk, and the ionization of the clumps, will be considered in the future works.

\begin{acknowledgements}
This work was supported by the National Natural Science Foundation of
China under grants 11333004, 11133005, 11573051, and U1531130, and the Fundamental Research Funds for the Central Universities under grants 20720150032.
FGX was supported in part by the Youth Innovation Promotion Association of CAS (id.\ 2016243), the National Basic Research Program of China (973 Program, grant 2014CB845800), the Strategic Priority Research Program `The Emergence of Cosmological Structures' of CAS (grant XDB09000000), and the CAS/SAFEA International Partnership Program for Creative Research Teams.
\end{acknowledgements}



\begin{thebibliography}{99}

\bibitem[2008]{BH08}
Balbus S. A., Henri P., 2008, \apj, 674, 408

\bibitem[2012]{BPN12}
Barai P., Proga D., Nagamine K., 2012, \mnras, 424, 728

\bibitem[2003]{BS03}
Blaes O., Socrates A., 2003, \apj, 596, 509

\bibitem[2001]{B01}
Begelman M. C., 2001, \apj, 551, 897

\bibitem[2014]{BA14}
Begelman M. C., Armitage P. J., 2014, \apj, 782, L18

\bibitem[2016]{Cao16}
Cao X. W., 2016, \apj, 817, 71

\bibitem[2010]{Chiang10}
Chiang C. Y., Done C., Still M., Godet O., 2010, \mnras, 403, 1102

\bibitem[2006]{Crummyetal06}
Crummy J., Fabian A. C., Ross R. R., 2006, \mnras, 365, 1067

\bibitem[2012]{DQ12}
Dexter J., Quataert E. 2012, \mnras, 426, L71

\bibitem[2007]{D07}
Done C., Gierli{\'n}ski M., Kubota A., 2007, \aapr, 15, 1

\bibitem[2010]{D10}
Done C., 2010, arXiv:1008.2287

\bibitem[1997]{E97}
Esin A. A., McClintock, J. E., Narayan R., 1997, \apj, 489, 865

\bibitem[1998]{G98}
Gammie C. F., 1998, \mnras, 297, 929

\bibitem[2008]{GDP08}
Gierli{\'n}ski M., Done C., Page K., 2008, \mnras, 388, 753

\bibitem[2010]{FR10}
Fabian A. C., Ross R. R., 2010, \ssr, 157, 167

\bibitem[1998]{K98}
Krolik J. H., 1998, \apj, 498, L13

\bibitem[1996]{K96}
Kuncic Z., Blackman E. G., Rees M. J., 1996, \mnras, 283, 1322

\bibitem[2005]{LMM05}
Liu B. F., Meyer F., Meyer-Hofmeister E., 2005, \aap, 442, 555

\bibitem[2007]{L07}
Liu B. F., Taam R. E., Meyer-Hofmeister E., Meyer F., 2007, \apj, 671, 695

\bibitem[2011]{L11}
Liu B. F., Done C., Taam R. E., 2011, \apj, 726, 10

\bibitem[2011]{LP77}
Liang E. P., Price R. H., 1977, \apj, 218, 247

\bibitem[2006]{Ma06}
Ma R. Y., Wang D. X., Zuo X. Q., 2006, \aap,  453, 1

\bibitem[2003]{MC03}
Maccarone T. J., Coppi P. S., 2003, \mnras, 338, 189

\bibitem[2002]{MC02}
Malzac J., Celotti A., 2002, \mnras, 335, 23


\bibitem[2006]{MMFR06}
Merloni A., Malzac J., Fabian A. C., Ross R. R., 2006, \mnras, 370, 1699

\bibitem[2005]{M05}
Meyer-Hofmeister E., Liu B. F., Meyer F., 2005, \aap, 432, 181

\bibitem[2007]{M07}
Meyer F., Liu B. F., Meyer-Hofmeister E., 2007, \aap, 463, 1

\bibitem[1994]{NY94}
Narayan R., Yi I., 1994, \apj, 428, L13

\bibitem[2008]{P08}
Petrucci P. O., Ferreira J., Henri G., Pelletier G., 2008, \mnras, 385, L88

\bibitem[1983]{P83}
Pozdnyakov L. A., Sobol I. M., Syunyaev R. A., 1983, Astrophysics and Space Physics Reviews, 2, 189

\bibitem[2012]{QL12}
Qiao E., Liu B. F., 2012, \apj, 744, 145

\bibitem[2017]{Sadowski17}
Sadowski A., Wielgus M., Narayan R., Abarca D., McKinney J. C., Chael A., 2017, \mnras, 466, 705

\bibitem[1973]{SS73}
Shakura N. I., Sunyaev R. A., 1973, \aap, 24, 337

\bibitem[1995]{ST95}
Shimura T., Takahara F., 1995, \apj, 445, 780

\bibitem[2002]{S02}
Smith D. M., Heindl W. A., Swank J. H., 2002, \apj, 569, 362

\bibitem[2012]{W12}
Wang J. M., Cheng C., Li Y. R., 2012, \apj, 748, 147

\bibitem[2016]{W16}
Wu M. C., Xie F. G., Yuan Y. F., Gan Z., 2016, \mnras, 459, 1543

\bibitem[2012]{X12}
Xie F. G., Yuan F., 2012, \mnras, 427, 1580

\bibitem[2016]{X16}
Xie F. G., Yuan F., 2016, \mnras, 456, 4377

\bibitem[2015]{Y15}
Yang Q. X., Xie F. G., Yuan F., Zdziarski A. A., Gierli{\'n}ski M., Ho L. C., Yu Z. L., 2015, \mnras, 447, 1692

\bibitem[2001]{Y01}
Yuan F., 2001, \mnras, 324, 119

\bibitem[2003]{Y03}
Yuan F., 2003, \apj, 594, L99

\bibitem[2014]{YN14}
Yuan F., Narayan R., 2014, \araa, 52, 529

\bibitem[2004]{ZG04}
Zdziarski A. A., Gierli{\'n}ski M., 2004, Progress of Theor. Phys. Supp., 155, 99

\bibitem[1997]{Zhang97}
Zhang S. N., Cui W., Chen W., 1997, \apj, 482, L155

\end{thebibliography}
\end{document}